\documentclass[useAMS,usenatbib]{mn2e}
\usepackage{amsfonts}
\usepackage{amsmath}
\usepackage{amssymb}

\usepackage[T1]{fontenc}

\usepackage[dvips]{graphicx}
\usepackage{myaasmacros}
\usepackage{color}
\usepackage{multirow}

\def\bea{\begin{eqnarray}}
\def\ena{\end{eqnarray}}

\newcommand{\RM}{\mathrm{RM}}
\newcommand{\RMin}{\mathrm{RRM}}
\newcommand{\RMinavg}{\left<\mathrm{RRM}\right>}
\newcommand{\aRMinavg}{\left<|\mathrm{RRM}|\right>}
\newcommand{\RMgal}{\mathrm{RM}_{\mathrm{gal}}}

\newcommand{\Sgal}{\sigma_{\mathrm{RM}}}
\newcommand{\Sgalavg}{\left<\sigma_{\mathrm{RM}}\right>}
\newcommand{\Lum}{L_{1.4\text{GHz}}}
\newcommand{\hp}{({\it hp})~}
\newcommand{\lp}{({\it lp})~}

\title[The RMs of high luminosity sources in the NVSS.]{The rotation measures of high luminosity sources as seen from the NVSS.}

\author[M. S. Pshirkov, P. G. Tinyakov, \& F. R. Urban]{M. S. Pshirkov$^{1,2,3}$\thanks{E-mail: pshirkov@sai.msu.ru}, P. G. Tinyakov$^{3,4}$\thanks{E-mail: petr.tiniakov@ulb.ac.be}, F. R. Urban$^4$\thanks{E-mail: furban@ulb.ac.be}\\
$^{1}$Sternberg Astronomical Institute, Lomonosov Moscow State University, Universitetsky prospekt 13, 119992, Moscow, Russia\\
$^{2}$Pushchino Radio Astronomy Observatory, 142290 Pushchino, Russia\\
$^{3}$Institute for Nuclear Research of the Russian Academy of Sciences, 117312, Moscow, Russia\\
$^{4}$Universit\'e Libre de Bruxelles, Service de Physique Th\'eorique, CP225, 1050, Brussels, Belgium\\
}

\begin{document}

\date{}

\pubyear{2015}

\maketitle

\label{firstpage}

\begin{abstract}
We re-analyse the subset of the Faraday rotation measures data from the NRAO VLA Sky Survey catalogue for which redshift and spectral index information is available, in order to better elucidate the relations between these observables.  We split this subset in two based on their radio luminosity, and find that higher power sources have a systematically higher residual rotation measure, once the regular field of the Milky Way is subtracted.  This rotation measure stands well above the variances due to the turbulent field of our Galaxy and measurement errors, contrarily to low power sources.  The effect is more pronounced as the energy threshold becomes more restrictive.  If the two sets are merged one observes an apparent evolution of rotation measure with redshift, but our analysis shows that this can be interpreted as an artifact of the different intrinsic properties of brighter sources that are typically observed at larger distances.
\end{abstract}


\section{Introduction}
\label{sec:intro}

Magnetic fields (MFs) seem to be omnipresent in the Universe, from the Earth to the huge intergalactic voids \citep{Kronberg:1993vk,Han:2002ns,Govoni:2004as,2004NewAR..48..763V,2012SSRv..166....1R}, including stars, galaxies, clusters, and perhaps filaments.  They were observed in galaxies at high redshift $z>1$ when the Universe was only a few billions years old \citep{Kronberg:2007dy,Bernet:2008qp}.

MFs also permeate the Large Scale Structure (LSS) of the Universe: it is
typically believed that they were initially created in the astrophysical
sources within the LSS, and only afterwards they polluted the LSS itself.
The MFs that are tentatively observed in the voids
\citep{Neronov:1900zz,Tavecchio:2010mk,Dolag:2010ni,Taylor:2011bn}\footnote{This
  is however by no means a settled issue, see
  \citep{Broderick:2011av,Murase:2011cy,Miniati:2012ge,Beck:2012cs,Neronov:2013zka,Saveliev:2013jda,Sironi:2013qfa}}
could also be blown away from the LSS; alternatively, they could be of
primordial origin---cosmological inflation, early universe phase transitions, etc
\citep{Grasso:2000wj,Dolgov:2003xd,Kandus:2010nw,Durrer:2013pga}.  Since there
are no compelling models for their genesis, it is crucial to better
understand their morphology, strength, spectral properties, and distribution
in the Universe; the more so for extragalactic fields, for which the very
large correlation lengths are theoretically difficult to achieve.

One of the most effective ways to study such extragalactic MFs is through the
observations of Faraday rotation measures (RMs).  The plane of polarization of a
linearly polarized electromagnetic wave of wavelength $\lambda$ travels
through a magnetized plasma rotates by the angle $\Delta \psi$ proportional to
the square of the wavelength,
\begin{equation}
\Delta\psi = \mathrm{RM}\cdot \lambda^2.
\label{RM1}
\end{equation}
Thus one needs multi- or at least bi-frequency observations in order to
determine the rotation measure RM.  The value of RM depends on the properties of the medium and the permeating magnetic field as follows, 
\begin{equation}
\mathrm{RM} = 812\int_{D}^{0} n_\text{e} B_{||} dl,
\label{RM2}
\end{equation}
where $n_e$ is the density of free electrons measured in cm$^{-3}$, $B_{||}$
is the component of the magnetic field parallel to the line-of-sight measured
in $\mu$G (positive when directed towards the observer), and $D$ is the
distance from the observer to the source in kpc.  Hence, an independent
estimate of the electron density $n_e$ is required to deduce
information on the magnetic field proper from Faraday rotation measures.

An indirect evidence of the MF presence in LSS and in voids may be obtained by 
studying the redshift evolution of RM of an ensemble of extragalactic sources. 
In a recent paper \citep{Neronov:2013lta} some evidence for a
significant redshift evolution in the RMs from the catalogue
\citep{Hammond:2012pn} was reported: the RMs 
were found to be growing with redshift in a way that could be interpreted as a
sign of non-zero nanoGauss-scale MFs in the filaments of the LSS.  Such redshift
dependence was not observed in the original catalogue \citep{Hammond:2012pn};
moreover, a recent work \citep{Banfield:2014roa} while re-examining the same
dataset (although retaining a smaller portion of if), did not find
indication for a very significant systematic correlation with redshift.  Finally, most recently
another analysis \citep{Xu:2014hya} reported on a quite weak evolution with
redshift, again for a very similar set.

We assess these claims in what follows, where in addition to previous analyses
we also search for a possible systematic dependence of the measured RMs on
intrinsic properties of sources, in particular their radio luminosity.  We
perform several tests, from which a coherent interpretation emerges:
\begin{itemize}
  \item we found no indication of a redshift evolution caused by the
    intervening medium;
  \item we do observe a sort of Malmquist
    bias, i.e., in a flux-limited sample we detect sources with higher luminosities at larger distances---the further we go, the higher the mean luminosity, thus
    mimicking an apparent redshift evolution by the redshift-dependent
    selection of sources with intrinsically different properties;
  \item the residual RM in low luminosity sources appears to be mostly due to the
    turbulent random Galactic MF (rGMF) and measurement errors, and consequently does not change with
    redshift;
  \item the residual RM in high luminosity sources instead shows a
    \emph{systematic} bias above the contribution from rGMF plus measurement errors; however, 
    there is also no clear redshift evolution in this set;
  \item this bias grows with more selective luminosity cutoffs, that is, there appears to be a positive correlation between the residual RM and the radio luminosity of the source.
\end{itemize}

The rest of this paper is organised as follows.  First, in
Sec.~\ref{sec:data_methods} we introduce the data and our selection, cleaning,
and averaging procedures.  Sec.~\ref{sec:results} reports all of our results
and their interpretation.  Finally, we summarise our findings in
Sec.~\ref{sec:summary}.

\section{Data and methods}
\label{sec:data_methods}

\paragraph*{The data.}  The largest set of RMs of extragalactic 
sources to date was compiled in \citep{Taylor:2009} from re-analyzing
the NRAO VLA Sky Survey (NVSS) data.  The NVSS is the largest by number survey of
polarized radio sources at declinations $>-40^{\circ}$
\citep{Condon:1998iy}.  The survey was performed in two nearby bands,
1364.9 and 1435.1 MHz; each having a width of 42 MHz.  Observations at
these close frequencies then give estimations of the RMs of the
sources.  The total number of observed sources was 37,543.  More than
10$\%$ of these sources have redshifts assigned
\citep{Hammond:2012pn}.

We selected 4002 NVSS sources with known redshifts from
\citep{Hammond:2012pn}.  Also we imposed the following cuts: to lower
the influence of the Galactic MF we accepted only sources with
$|b|>20^{\circ}$, and we dismissed all the sources with
$|\mathrm{RM}|>300~\mathrm{rad~m^{-2}}$ owing to the fact that RMs obtained in
two close frequencies are not fully reliable if their absolute values
are too large \citep{Taylor:2009}.  That left us with 3647 sources.

\paragraph*{Removing the GMF.}  
Each observed $\RM$ is the sum of several contributions: 
the one due to the regular GMF which we denote $\RMgal$, 
the one due to rGMF, the RM intrinsic to the source, and finally the rotation
acquired while travelling through the intergalactic medium. 

Due to their random character, the last three contributions cannot be separated on a source-by-source basis.  On the contrary, $\RMgal$ can, in principle, be estimated
and subtracted for each source:\\
\[
\RMin=\RM-\RMgal.
\]
where we introduced the residual RM ($\RMin$).  Clearly, any redshift evolution or correlation with luminosity would have a more pronounced effect on $\RMin$ than on $\RM$.

The Galactic contribution $\RMgal$ was estimated using the observed RMs themselves.  In order to do so, we first cleaned the {\em full} NVSS RM catalogue removing the outliers following the algorithm described in \citep{Pshirkov:2013wka} (a similar approach was used in \citep{Xu:2014hya}, while an alternative algorithm has been devised and applied in \citep{Oppermann:2011td,Oppermann:2014cua}: our results are unchanged if we employ their compilation; notice also that ionospheric RM variation is negligible: \citep{Sotomayor-Beltran:2013vma}).  This algorithm is very simple: a circle of $3^{\circ}$ radius was circumscribed around every source in the catalogue and both the average RM and its variance
were calculated for the selected region. If the RM of the source was more than
two r.m.s.\ values away from the average, the source was marked as
``outlier''.  In total, 1974 sources were removed after this procedure,
leaving 35,569 in the clean set.

Then, for each source with an assigned redshift, we averaged
the RMs from the cleaned catalogue within the $3^{\circ}$ circle
around the source (typically, about 30 values). We interpreted
the average as $\RMgal$ corresponding to that source.  Within the same circle, we also calculated the standard deviation $\Sgal$, which measures the dispersion of RM due to the rGMF and other factors.  The contributions to $\Sgal$ are assumed to be simple Gaussian errors on each individual source, including measurement errors.

\paragraph*{Luminosity.}
We employed the spectral indices reported in the recent work \citep{Farnes:2014a}: we first identified as many sources as possible from the \citep{Hammond:2012pn} set, and assigned them their respective $\alpha$.  We were able to do so for overall 3051 sources out of 3647.

Once the spectral index was assigned to as many sources as possible,
we calculated the luminosity with the help of the relation
\begin{equation}\label{lumi}
\Lum = \frac{4\pi D_L^2 S_{1.4\text{GHz}}}{(1+z)^{\alpha+1}},
\end{equation}
where $D_L$ is the luminosity distance and $S_{1.4\text{GHz}}$ 
the flux density at 1.4 GHz \citep{Hogg:1999ad}.  In calculating the luminosity distance $D_L$ we chose a $\Omega_m=0.73$, $\Omega_\Lambda=0.27$, and $H=71\text{km}/\text{s}/\text{Mpc}$.

In the Appendix we show that our results do not change if we employ our own idependent compilation of spectral indices.

\section{Results}
\label{sec:results}

With the final set at hand, we binned all sources in redshift.  The criteria with which we chose the bins are explained below. We define the averages $\RMinavg \equiv \sum_\text{bin}\RMin/N$, with $N$ the number of sources in each bin, that is, the mean $\RMin$ in each (redshift, latitude) bin. We work with the means (rather than, say, variances) because they are believed to be more sensitive to the presence of intergalactic magnetic fields, see \citep{Blasi:1999hu}. Similarly, $\Sgalavg \equiv \sum_\text{bin}\Sgal/N$.

We will use $\Sgalavg$ to estimate the combined contribution of the rGMF, source-intrinsic RM, measurement errors, and the intergalactic magnetic field on the $\RMin$ of our target sources. We assume that these contributions to $\RMin$ have very similar statistical distributions for nearby lines of sight; in that case the $\Sgal$ that we calculate from surrounding sources also measures the spread in $\RMin$ that we would expect for each target source if its $\RMin$ follows the same distribution as the $\RMin$ of surrounding targets.

Notice that here, as well as everywhere else in the paper, the errors are given by $\sigma(|\RMin|)_\text{bin} / \sqrt{N}$ (and similarly for $\Sgal$: $\sigma(\Sgal)_\text{bin} / \sqrt{N}$), where $\sigma(X)$ is the standard deviation of $X$, and $N$ the number of sources in each bin.

\begin{figure}
\begin{center}
  \includegraphics[width=0.48\textwidth]{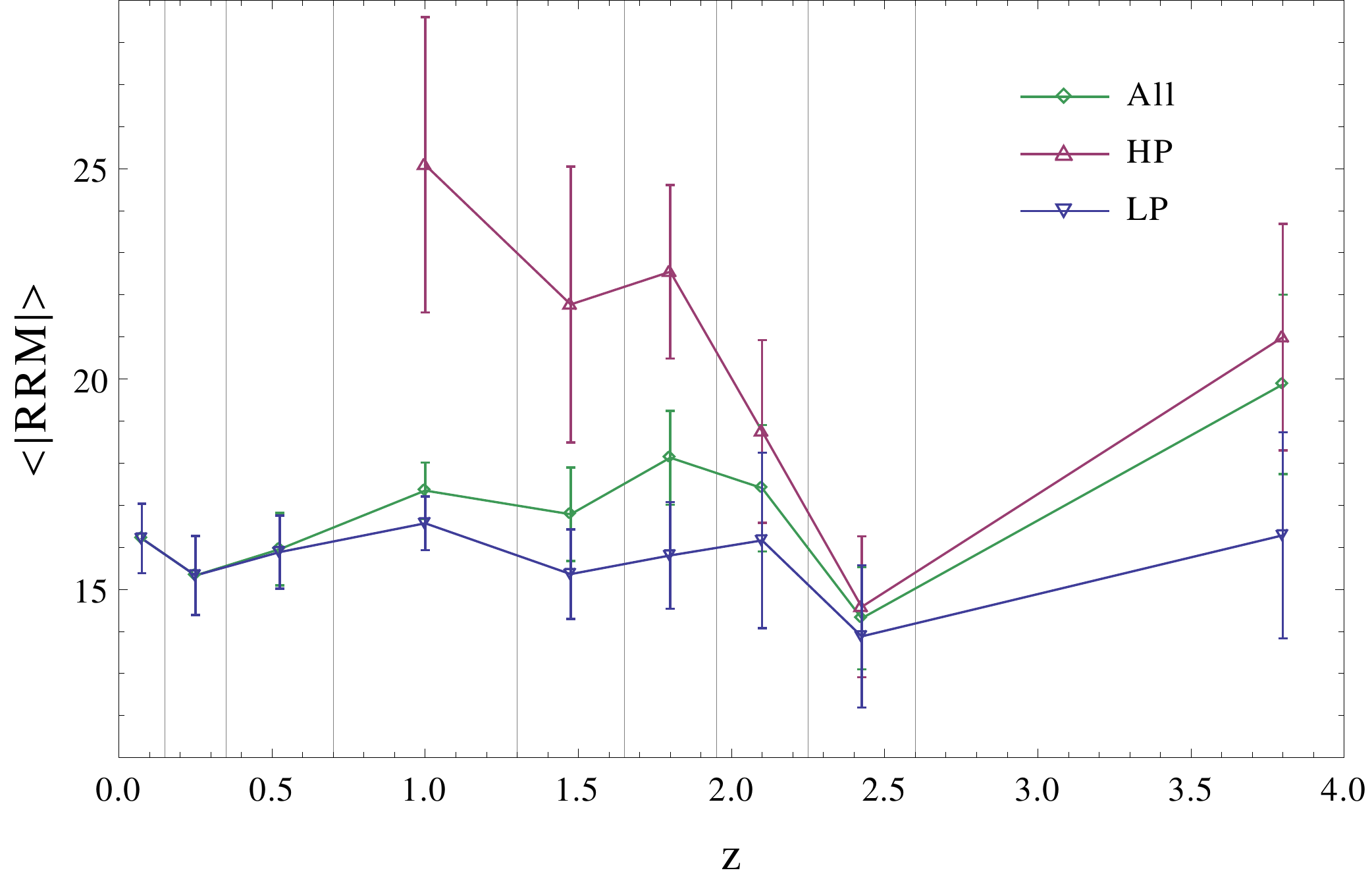}
\end{center}
\caption{Redshift-dependence of $\aRMinavg$ for the entire set, and for the \lp and \hp sets separately.  All quantities are expressed in $\text{rad}/\text{m}^2$.}
\label{fig:mean20j}
\end{figure}

One last note about notation: in all our figures where rotation measures appear, it is intended that their units are always $\text{rad}/\text{m}^2$---we omit these in the figures to avoid cluttering them.  Fig.~\ref{fig:mean20j} (green diamonds) shows that there might exist an apparent redshift evolution of $\aRMinavg$; in fact, our results appear in
good qualitative agreement with the results of \citep{Neronov:2013lta}. The quantitative difference could arise due to the different procedures of foreground (i.e., GMF) subtraction.

However, this agreement disappears after we further split the set of sources into two subsets according to their intrinsic radio luminosity. Placing the luminosity cut-off at $L = 10^{27.8} \mathrm{WHz}^{-1}$ gives 2593 sources in the low-power \lp group, and 457 sources in the high-power \hp group. This luminosity cut-off allows us to use six bins at high redshift and high radio luminosity. We tried several combinations of redshift bins and luminosity cut-offs, and settled for one which gives us a sufficiently large sample of sources in each bin in Galactic latitude or redshift, as well as enough resolution to check for trends with Galactic latitude or redshift. The limited number of available sources makes it impossible to split the data even further into, for example, sources with different physical properties.

The binning procedure was as follows.
\begin{itemize}
  \item We first looked at the high power set.  Going from the highest
    $z$, we chose the bins such that there are about 80 \hp
    sources in each bin. This gave us 6 bins starting at redshift
    $z=0.7$ and upward. 
  \item At redshifts below $0.7$ there are only 5 sources in the ({\it
    hp}) sample. For this reason, at low redshifts the binning was set
    according to the \lp sample. We decided to keep three
    additional low-redshift bins and again chose the bin boundaries
    so as to have about the same number of events in each bin.  
\end{itemize}
The resulting mean redshifts of the bins $z_m$---starting from $z=0$, as well as the numbers of \lp and
\hp sources in each bin are summarised in Table~\ref{tab:bins}.

\begin{center}
\begin{table}
\resizebox{0.45\textwidth}{!} {
\begin{tabular}[ht]{|c||*{9}{c}|}
  $z_m$ & 0.075 & 0.25 & 0.525 & 1.0 & 1.475 & 1.8 & 2.1 & 2.425 & 3.8 \\ \hline
  \lp & 418 & 418 & 501 & 677 & 291 & 137 & 76 & 50 & 25 \\ \hline
  \hp & 0 & 0 & 5 & 68 & 83 & 72 & 70 & 79 & 80 \\ \hline
\end{tabular}}\caption{Number of sources in the \lp and \hp sets in each redshift bin with mean redshift $z_m$.  Notice that the overall count is 3050 (not 3051), since one source has $z>5$.}\label{tab:bins}
\end{table}
\end{center}

The upward and downward triangle points in Fig.~\ref{fig:mean20j} show the impact of the separation into the
\lp and \hp sets. There seems to be a systematic
shift in $\aRMinavg$ between the two sets; the shift is
not very large (on the order of $5~\mathrm{rad~m^{-2}}$) but is
coherent throughout most of the bins. At the same time, neither \lp
nor \hp separately show a systematic dependence of $\aRMinavg$ with $z$.

The most straightforward interpretation of this first result is that:
(a) there is no significant evolution with redshift, and (b) higher
$\aRMinavg$ correlate with higher power.

The featureless binning in $z$ for the \lp set seems also to be at odds with the model proposed in \citep{Beck2013}.  The final result of the galactic dynamics outlined in that work is that it is not uncommon for host galaxies to possess extended and strongly magnetised halos, which result in a (truly) intrinsic RM around $1000~\mathrm{rad~m^{-2}}$ already at $2<z<4$; if this type of galaxy represented a significant part of our sample then $\aRMinavg$ would increase up to $300~\mathrm{rad~m^{-2}}$ at these redshifts, the cutoff in rotation masures that we imposed in this work: this is not the case in our analysis.

\begin{figure}
\begin{center}
\includegraphics[width=0.48\textwidth]{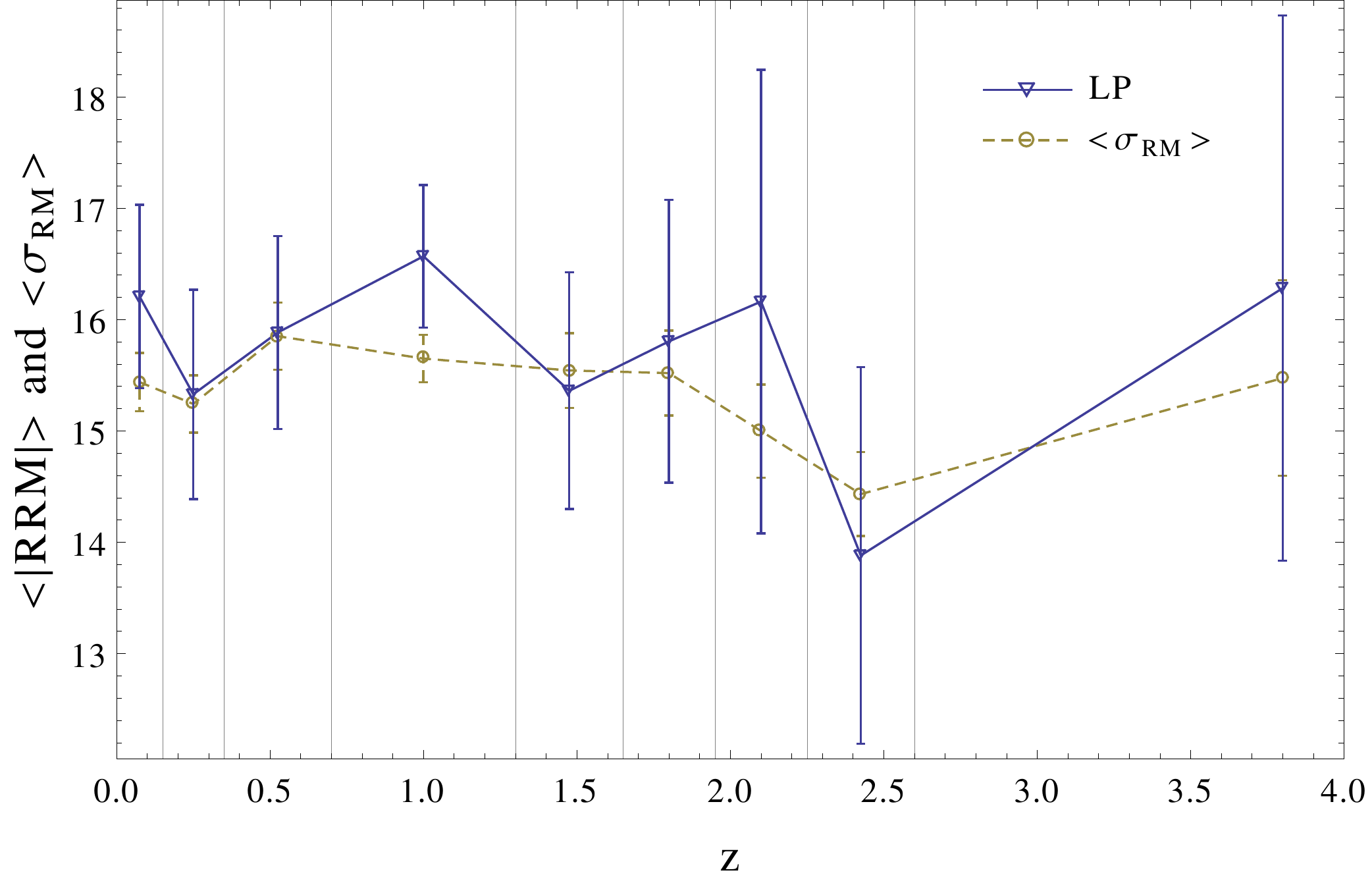}\\\vspace*{5pt}
\includegraphics[width=0.48\textwidth]{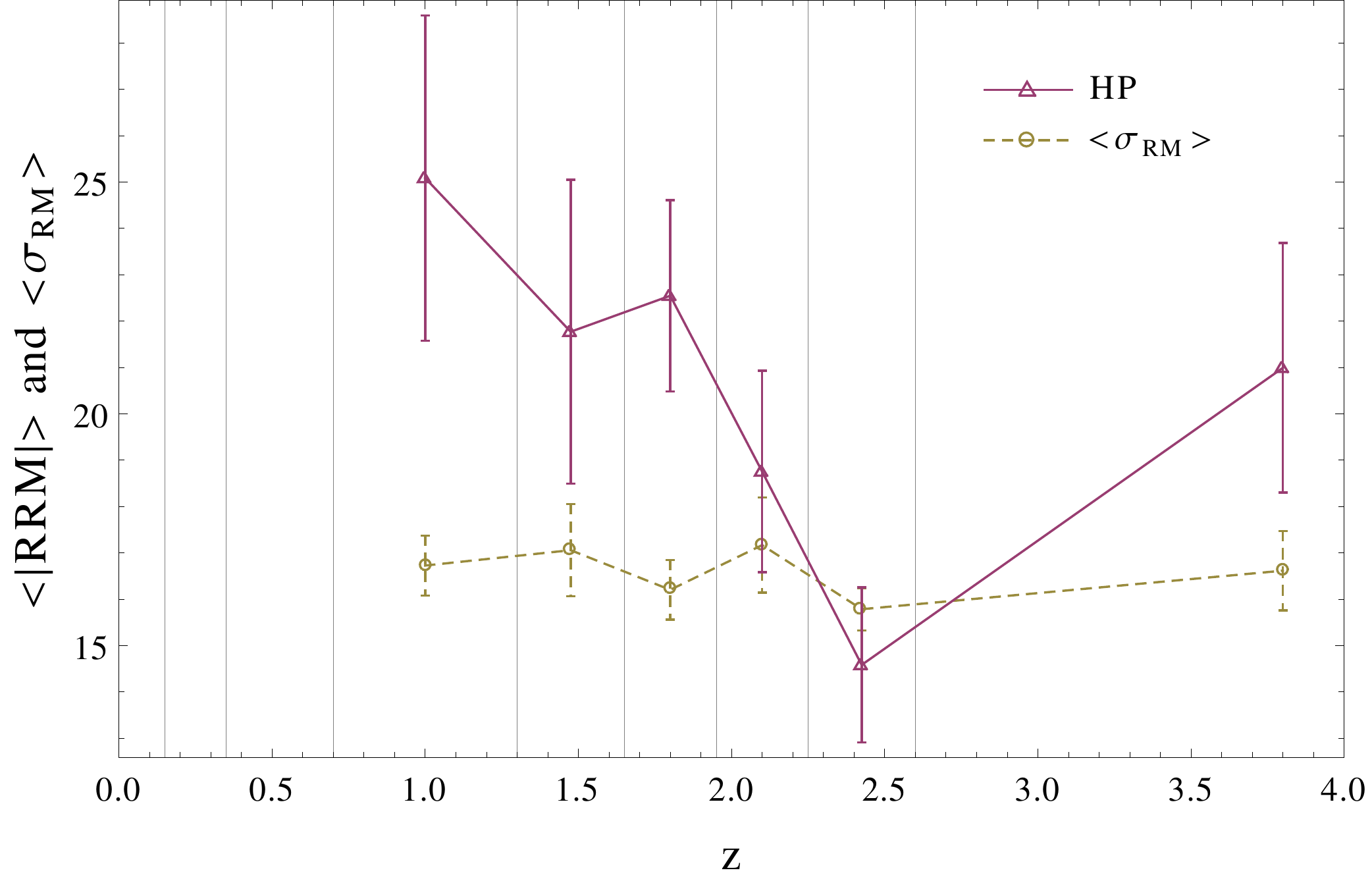}
\end{center}
\caption{Redshift dependence of $\aRMinavg$ and $\Sgalavg$ in the 
\lp (upper panel) and \hp (lower panel) groups.  All quantities are expressed in $\text{rad}/\text{m}^2$.}\label{fig:SZ}
\end{figure}

Motivated by this initial result, we have performed several tests in order to assess the validity of this conclusion.  Fig.~\ref{fig:SZ} shows again
$\aRMinavg$ now together with $\Sgalavg$ for the \lp and \hp sets.  There is nearly no difference in the $\Sgalavg$ for the two sets, as could be expected because the $\Sgalavg$ are calculated from the full NVSS RM catalogue. Second, there is no obvious trend with redshift for the variances. 



\begin{figure}
\begin{center}
\includegraphics[width=0.48\textwidth]{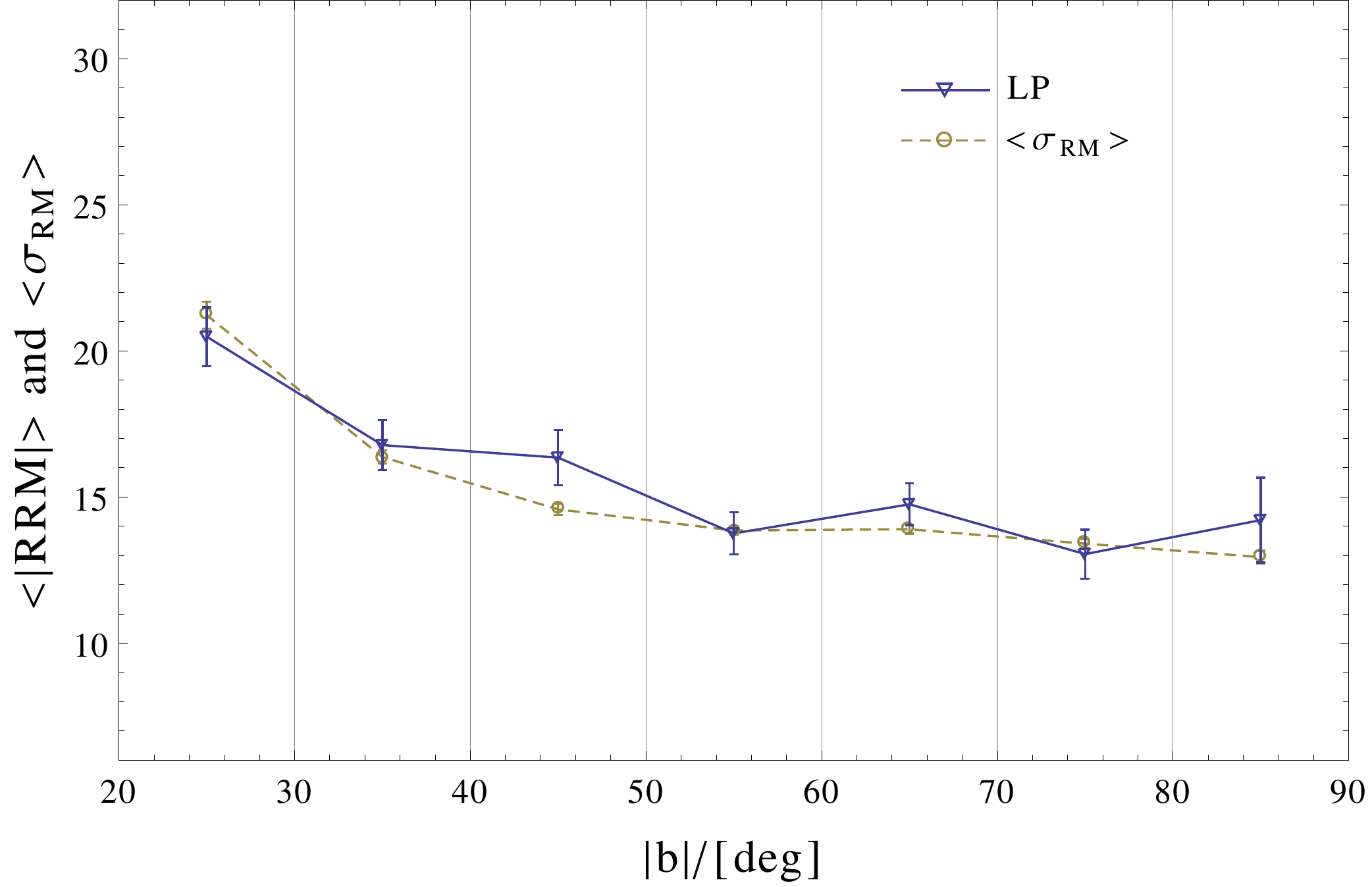}\\\vspace*{5pt}
\includegraphics[width=0.48\textwidth]{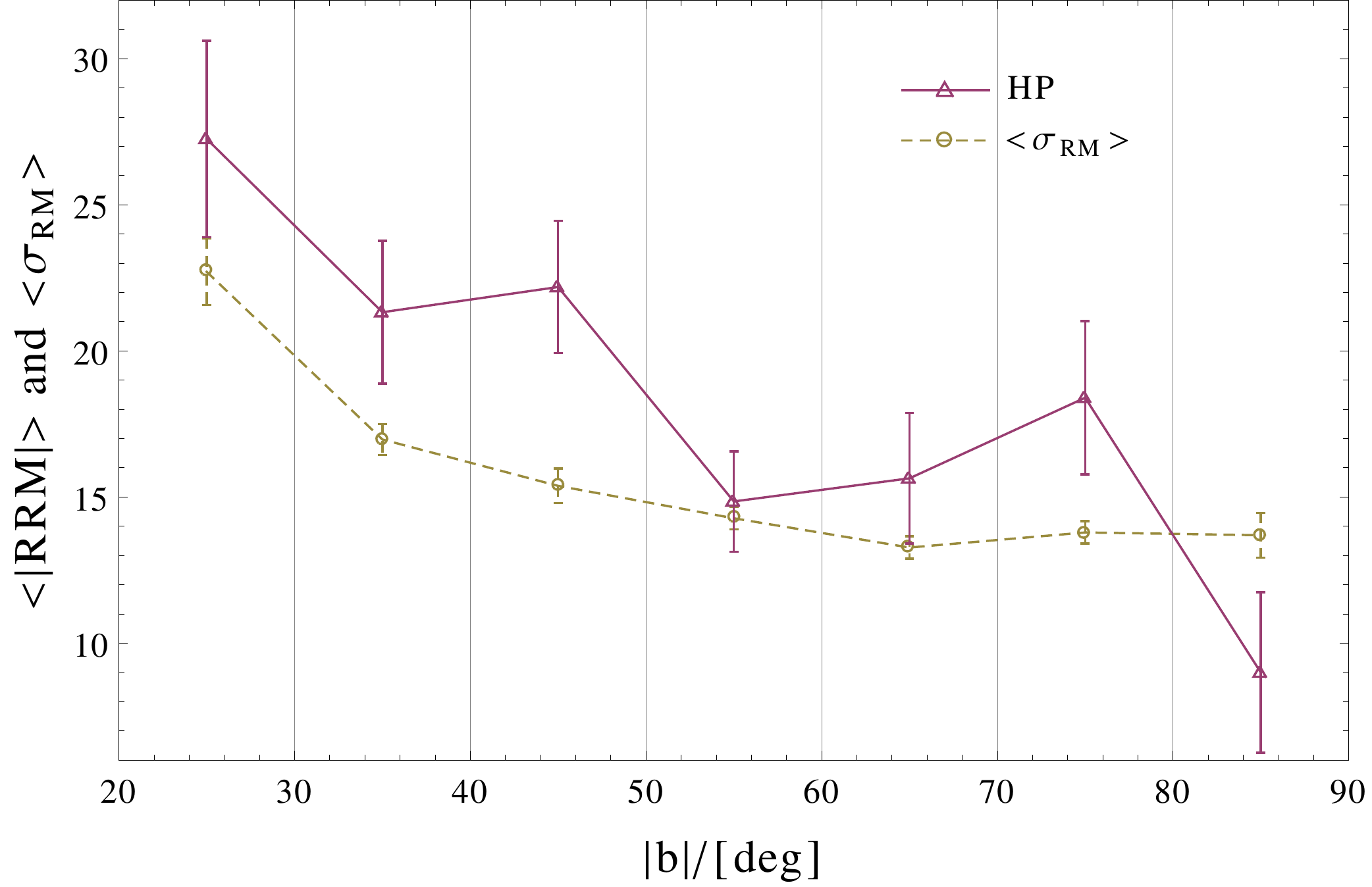}
\end{center}
\caption{Latitude dependence of $\aRMinavg$ and $\Sgalavg$ in the \lp (upper panel) and \hp (lower panel) groups.  All quantities are expressed in $\text{rad}/\text{m}^2$.}\label{fig:SB}
\end{figure}

In Fig.~\ref{fig:SB} we have binned the data in Galactic latitude instead of redshift. $\Sgalavg$ shows a very clear dependence on latitude $b$, which can be modelled by a latitude-dependent component plus a constant value of $13~\mathrm{rad~m^{-2}}$ \citep{Pshirkov:2013wka}. This is nicely compatible, numerically as well as qualitatively, with the results of \citep{Schnitzeler:2010ax}, where the latitude dependence of the variance was pointed out, and the different contributions were identified. The $\aRMinavg$ of data points in the \lp category are compatible with $\Sgalavg$, but the $\aRMinavg$ of the \hp ones present a systematic coherent shift over $\Sgalavg$ by about $5~\mathrm{rad~m^{-2}}$.

It is tempting to interpret the absence of redshift evolution in the \lp group as most likely coming from the turbulent magnetic field of the Milky Way itself, but also errors associated with each RM measurement contribute significantly to the $\RMin$ of the \lp sources.


There is the possibility that the features observed when marginalising over one variable (latitude $b$ or redshift $z$) could pollute the other ``alternative'' marginalisation.  We did check for this possibility by only choosing higher latitudes, where $\Sgalavg$ is constant, or different redshift bin sizes, and we did not observe any significant departure from the conclusions we arrive at.

In principle a possible explanation for the different behaviour of the \hp group could be that the sources in this group are distributed differently, i.e. closer to the Galactic plane, and that would lead to the observed excess.  In fact, there is a slight preference for low-$b$ in the \hp set; however, as we demonstrated, it is impossible to attribute all our excess to this small ``bias'' when we binned with latitude itself.  This explanation is therefore ruled out.  That means that the positive correlation between the $|\RMin|$ and
radio luminosity is real, and seemingly compatible with its arising close to the source.

\begin{figure}
\begin{center}
\includegraphics[width=0.48\textwidth]{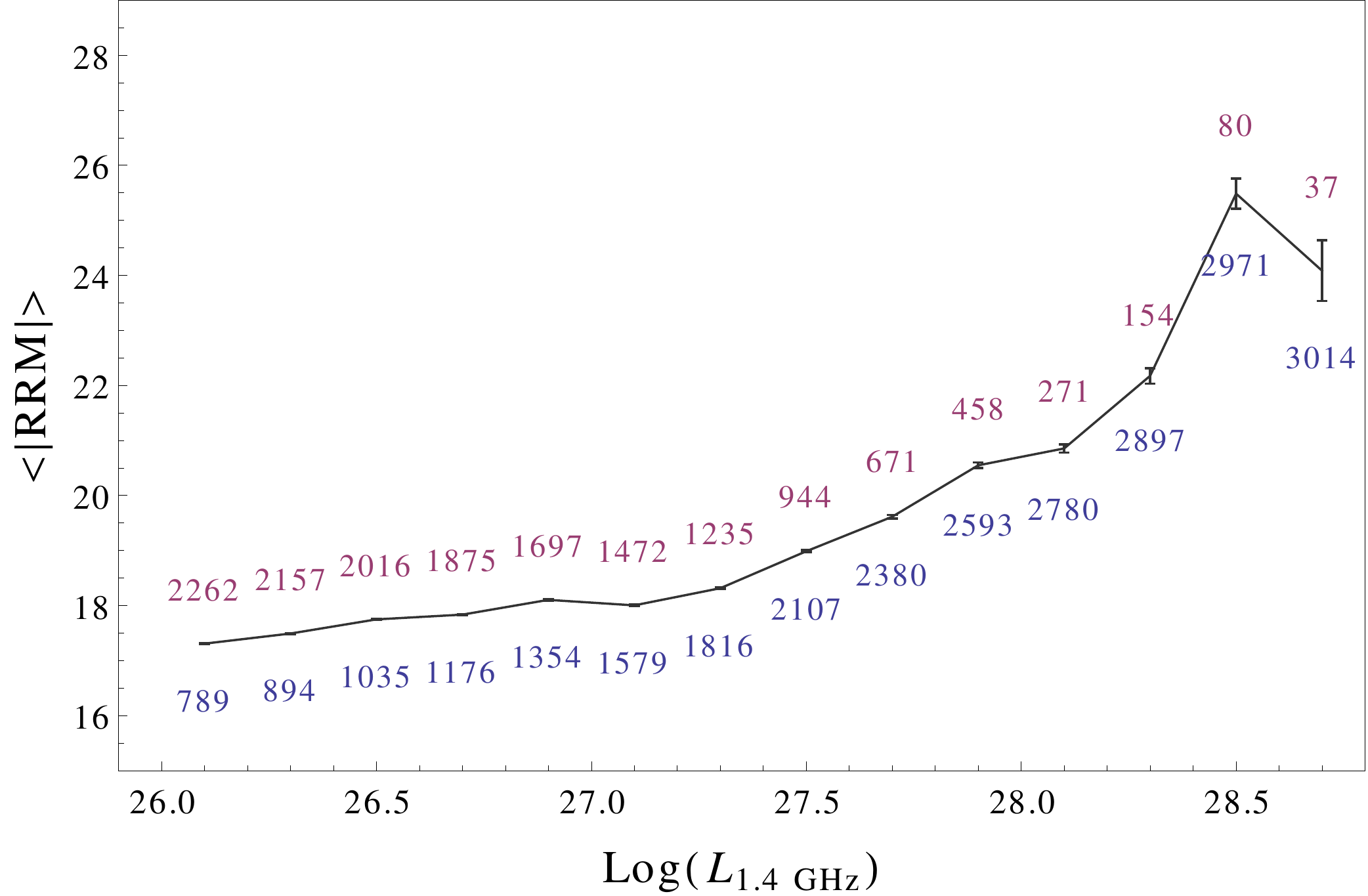}
\end{center}
\caption{Overall mean $|\RMin|$ for the \hp set plotted 
against luminosity cut.  The values above and below the curve are 
the number of events at given power threshold in the \hp 
and \lp sets, respectively, as defined by the cut.  All quantities are expressed in $\text{rad}/\text{m}^2$.}\label{fig:hzhp}
\end{figure}

We now turn our attention to the effect of the luminosity cutoff, to check for
a possible dependence on the particular value we have chosen.  We begin with
Fig.~\ref{fig:hzhp}: we show here the double averaged $|\RMin|$ (average over all targets in each latitude bin, then averaged over all latitude bins) of all sources in
the \hp set, as a function of the radio luminosity where we split the
entire set in two.  The trend towards a more pronounced $|\RMin|$ with more
severe threshold is very clear.  This shows how the threshold itself is not
important, and that in fact if we were to choose a cut-off at higher luminosity, were we
not limited by statistics, the results would be even more significant. We reach the same conclusion if we only include sources above a certain redshift. Therefore, distance-related selection effects are not important in our analysis.

\begin{figure}
\begin{center}
\includegraphics[width=0.48\textwidth]{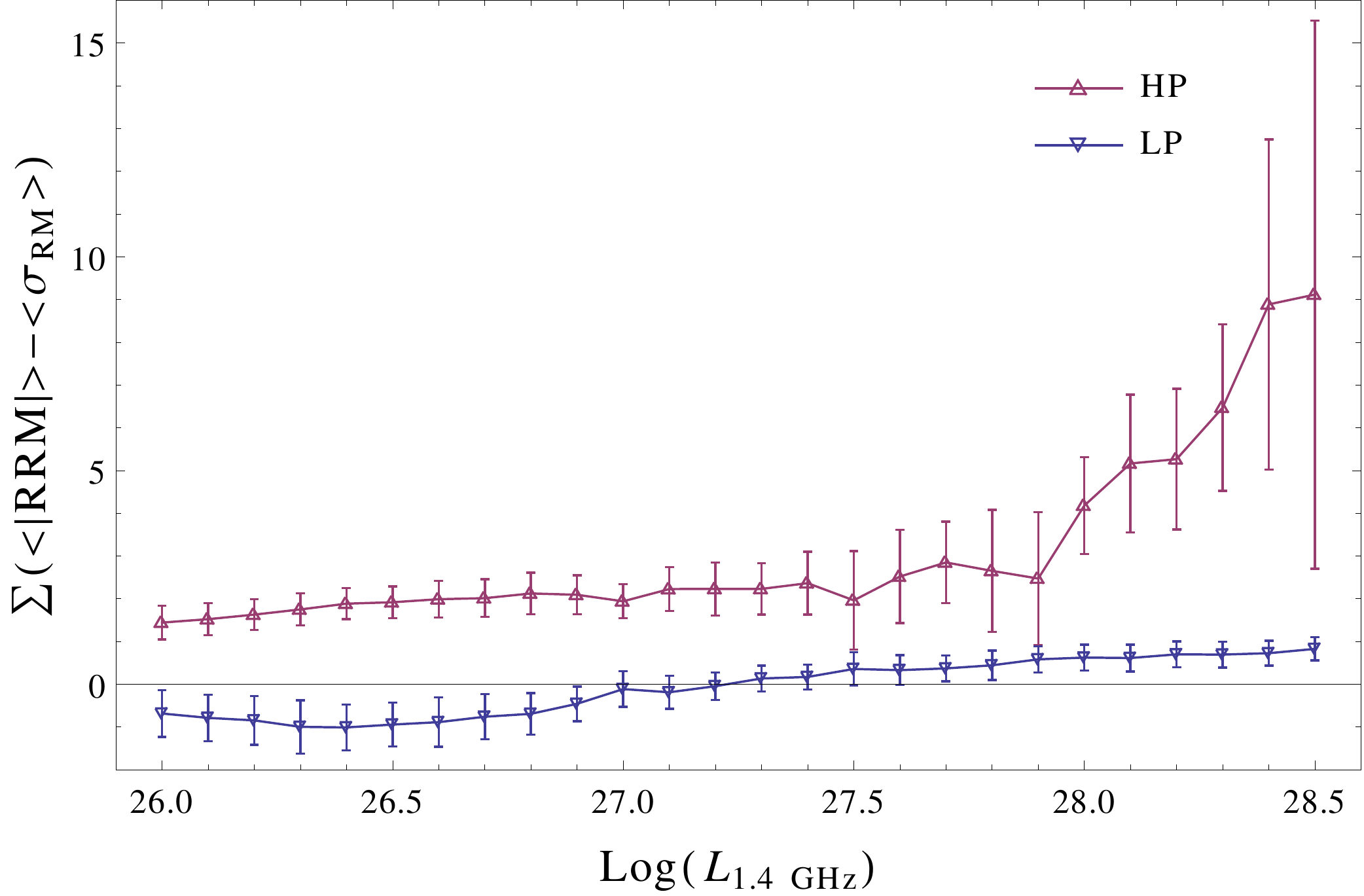}
\end{center}
\caption{Sum of the differences $\aRMinavg - \Sgalavg$ for the \lp and \hp sets in each latitude bin, followed by summing all latitude bins,
plotted against the power cutoff threshold.  All quantities are expressed in $\text{rad}/\text{m}^2$.}\label{fig:stack}
\end{figure}

To include also the change in $\Sgalavg$ as a function of Galactic latitude in our discussion of Fig.~\ref{fig:hzhp}, we calculate the difference between $\aRMinavg$ and $\Sgalavg$ for each latitude bin from Fig.~\ref{fig:SB}, then sum all latitude bins.  This was done for varying luminosity cutoffs, similarly to Fig.~\ref{fig:hzhp}.  The results are presented in Fig.~\ref{fig:stack}.  Independent of our choice for the luminosity cut-off, the $\aRMinavg$ of sources from the \lp set are consistent with $\Sgalavg$. On the other hand, the $\aRMinavg$ of sources from the \hp set are incompatible with $\Sgalavg$ for all luminosity cut-offs. This strengthens the conclusion that we reached from analysing Fig.~\ref{fig:SB}, which was limited to a single luminosity cut-off. Bear in mind that only the difference in behaviour between the \lp and \hp sources matters here, not the actual values of $\aRMinavg -\Sgalavg$ (which in fact are not predictable individually).


Having established that there seems to be a positive correlation between
$\RMin$ and luminosity, a legitimate question is whether a relatively small
group of sources with (perhaps) extreme properties is driving this result---for instance, there could be different populations of sources with different intrinsic properties.
Removing sources from the \hp set with large $|\RMin|$ still produces an offset, but the error bars increase, and it is no longer clear if these $\aRMinavg$ are incompatible with $\Sgalavg$ in this case. Also, employing the spectral index $\alpha$ as discriminant, we were not able to clearly discern between two (or more) populations, and we did not observe any particular trend of the spectral index $\alpha$ with $\RMin$ itself.


Yet another possibility, put forward in the works of \citep{Joshi:2013oka,Farnes:2014b,Banfield:2014roa}, is that the effect could be caused not by intervening filaments but by smaller systems like MgII absorbers.  Thence, sources with MgII absorption along the line of sight have higher RMs than those without absorption.  Since sources at higher redshifts (and hence preferentially higher luminosities) have more absorbers than sources at low redshifts (where although high luminosities are present, on average their luminosities will be lower), the correlation we observe might be a result of the intervening systems rather than due to luminosity itself.  If this were true, however, we would see this in both \lp and \hp sets, but the former does not show any such effect (although the fluctuations due to measurement errors and the random GMF can be large).

\section{Summary}\label{sec:summary}

To conclude, we briefly recapitulate the salient features of our searches.  We set out with the purpose of investigating the possibility of a redshift dependence in the observed Faraday sky, investigation which we based on the set of all NVSS catalogue sources for which redshift information is known---this is the largest available set in the literature at the moment.  The catalogue was cleaned removing outliers with potentially unreliable RMs; we then used the data itself to separate the RM due to the regular MF of the Milky Way: all our statistical results are based on the \emph{residual} RM ($|\RMin|$).

We specifically looked for the effect of the radio luminosity $\Lum$ of the sources (calculated independently for most sources through their flux densities); this effect and our interpretation of our results can be summarised in these seven points below.
\begin{itemize}
  \item The $|\RMin|$ positively correlates with $\Lum$, that is, the higher luminosity sources have higher residual RMs (Fig.~\ref{fig:hzhp}).
  \item The $|\RMin|$ of low luminosity sources is dominated by the variance due to measurement errors and that coming from the rGMF (Figs.~\ref{fig:SZ} and~\ref{fig:SB}, top).  The overall $|\RMin|$ consistently decreases with latitude as we move away from the Galactic plane.
  \item The $|\RMin|$ of high power sources, on the other hand, stands out coherently above the expected $\Sgal$ variance; this is true in both redshift and latitude bins, where in the latter it is also easy to single out the Galactic contribution (Figs.~\ref{fig:SZ} and~\ref{fig:SB}, bottom).
  \item Therefore, there is an \emph{overall} shift in $|\RMin|$ between the two sets.  This shift appears in redshift bins (Figs.~\ref{fig:mean20j} and~\ref{fig:SZ}) as well as in latitude bins (again Fig.~\ref{fig:SB}), and amounts to the same value of approximately $5~\mathrm{rad~m^{-2}}$ for a split at $\log \Lum = 27.8$.
  \item The systematic shift is not an artifact of the luminosity cutoff, as it actually grows with more constraining choices (Figs.~\ref{fig:hzhp} and~\ref{fig:stack}).
  \item The systematic shift is also not an artifact of redshift evolution, as we do not observe any trend in the $z$ behaviour of either set (Figs.~\ref{fig:mean20j} and \ref{fig:SZ}).
  \item If we ignore the luminosity and analyse the full catalogue we do observe a weak redshift dependence (Fig.~\ref{fig:mean20j}), which we can hence impute to a Malmquist bias, i.e., from larger distances brighter sources are more easily detected.  Again, what does appear to correlate are $|\RMin|$ and luminosity, not redshift.
\end{itemize}

These results are promising, and it would be extremely useful to understand if this correlation is physical and isolate its origin: we performed a few tests in this sense but the statistical size of the sample was too limiting a factor.  In particular it would be very interesting to see whether the correlation and/or the systematic shift are driven by a particular set of sources, for instance a small set belonging to a particular type of objects.  With a larger dataset these questions could be easily addressed; a larger set would also allow a much better estimation of the Galactic contribution, and would finally shed light on possible features as for any redshift development of $|\RMin|$.  We leave all these updates for future investigations.

\section*{Acknowledgements}

The authors would like to thank B Gaensler and J L Han for positive feedback and useful correspondence. MP wishes to thank the Service de Physique Th\'eorique in Brussels, where this work was conceived.  The work of MP is supported by RFBR Grants No.~13-02-01293a, by the Grant of the President of Russian Federation MK-2138.2013.2 and by the Dynasty Foundation.  FU and PT are supported by IISN project No.~4.4502.13 and Belgian Science Policy under IAP VII/37. PT is supported in part by the RFBR grant 13-02-12175-ofi-m.  This research has made use of NASA's Astrophysics Data System.

\section*{Appendix}

\begin{figure}
\begin{center}
\includegraphics[width=0.48\textwidth]{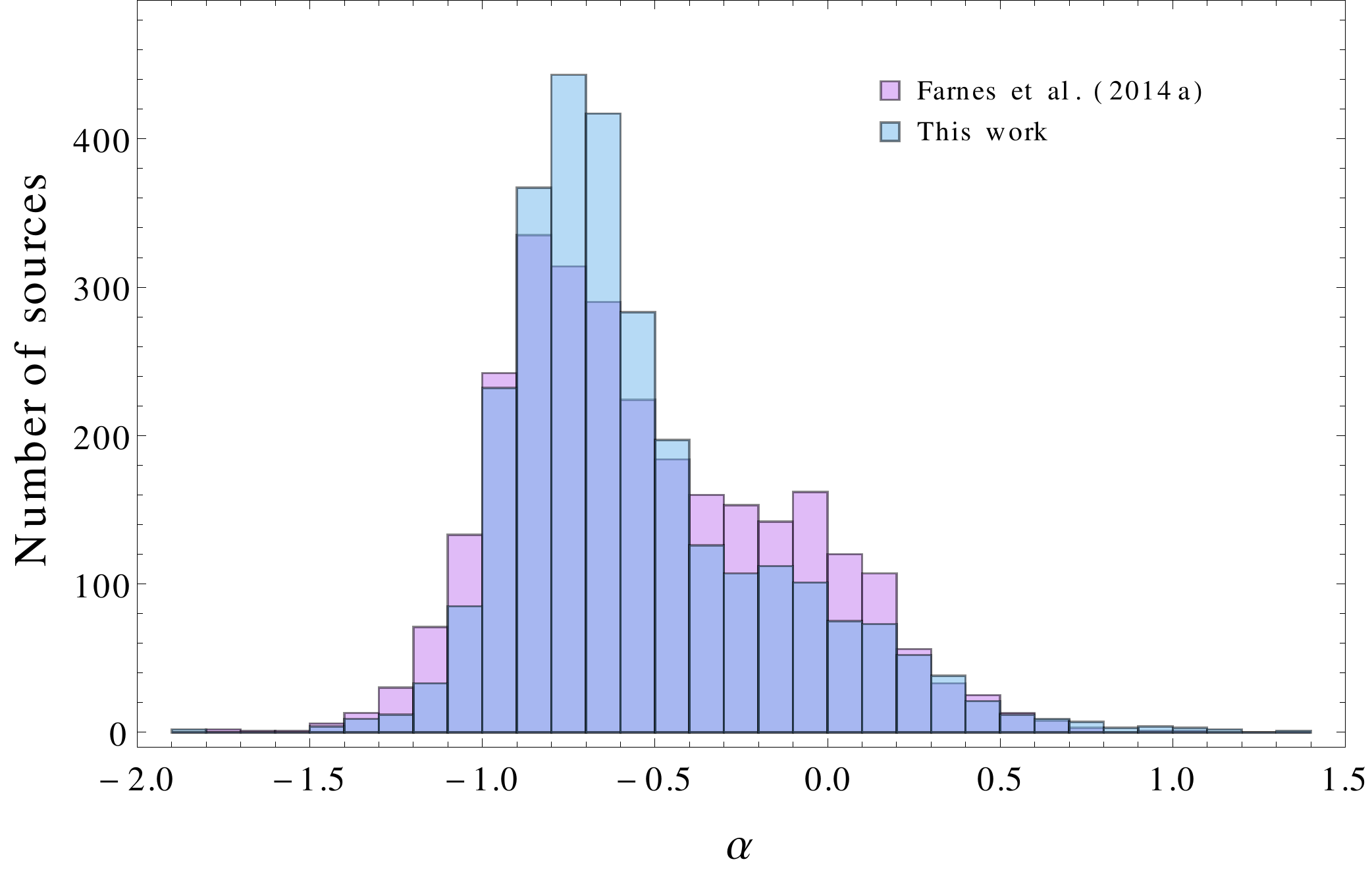}
\end{center}
\caption{The different $\alpha$ distributions for the common 2830 sources of \citep{Farnes:2014a} and those identified with our own compilation of VLSS, WENSS, SUMSS, and VcV.}\label{fig:alphas}
\end{figure}

One crucial ingredient in this analysis is the intrinsic luminosity of the sources.  In addition to using the compilation of \citep{Farnes:2014a}, we have manually computed the spectral indices of as many sources as possible using information from three additional catalogues: VLSS\footnote{VLA Low-Frequency Sky Survey \citep{2007AJ....134.1245C}.} (74 MHz, $\text{dec} >-30^{\circ}$), WENSS\footnote{The Westerbork Northern Sky Survey \citep{1997AAS..124..259R} http://www.astron.nl/wow/testcode.php?survey=1\&more=1.} (352 MHz, $\text{dec} >28.5^{\circ}$), and SUMSS\footnote{The Sydney University Molonglo Sky Survey \citep{1999AJ....117.1578B,2003MNRAS.342.1117M}.} (843 MHz, $\text{dec} <-30^{\circ}$), from which we can calculate the spectral index $\alpha$ of the source.  When combined, they nicely cover all the sky.  The algorithm to obtain the intrinsic power of each source was as follows:
\begin{enumerate}
  \item for each item in the catalogue of \citep{Hammond:2012pn} (with
    additional cuts at $|b|>20^{\circ}$ and
    $|\mathrm{RM}|<300~\mathrm{rad~m^{-2}}$) we found the counterpart in one or
    more of the three catalogues mentioned above;
  \item wherever possible, that is, where at least two different
    fluxes are available, we calculated $\alpha$;
  \item if $\alpha$ could be calculated from either SUMSS or WENSS 
    we employed these values, because all VLSS entries have larger error;
  \item conversely, if only data from VLSS was available, we used the latter;
  \item only as a last option we calculated $\alpha$ from the 
    \citep{2010AA...518A..10V} compilation as it is comparatively less reliable.
\end{enumerate}
With this procedure, we obtained the final set of 3190 sources 
(out of the original 3647) for which $\alpha$ was assigned, while with \citep{Farnes:2014a} we were able to work with 3050 sources.  In Fig.~\ref{fig:alphas} we show the different distributions for the sources which belong to both catalogues.

\begin{center}
\begin{table}
\resizebox{0.45\textwidth}{!} {
\begin{tabular}[ht]{|c||*{9}{c}|}
  $z_m$ & 0.075 & 0.25 & 0.525 & 1.0 & 1.475 & 1.8 & 2.1 & 2.425 & 3.8 \\ \hline
  \lp & 474 & 450 & 528 & 693 & 289 & 135 & 74 & 51 & 27 \\ \hline
  \hp & 0 & 0 & 5 & 74 & 81 & 76 & 77 & 78 & 78 \\ \hline
\end{tabular}}\caption{Number of sources in the \lp and \hp sets in each redshift bin for our set of sources with manually calculated luminosity.  The overall count is of 3190 objects.}\label{tab:binsOLD}
\end{table}
\end{center}

\begin{figure}
\begin{center}
\includegraphics[width=0.48\textwidth]{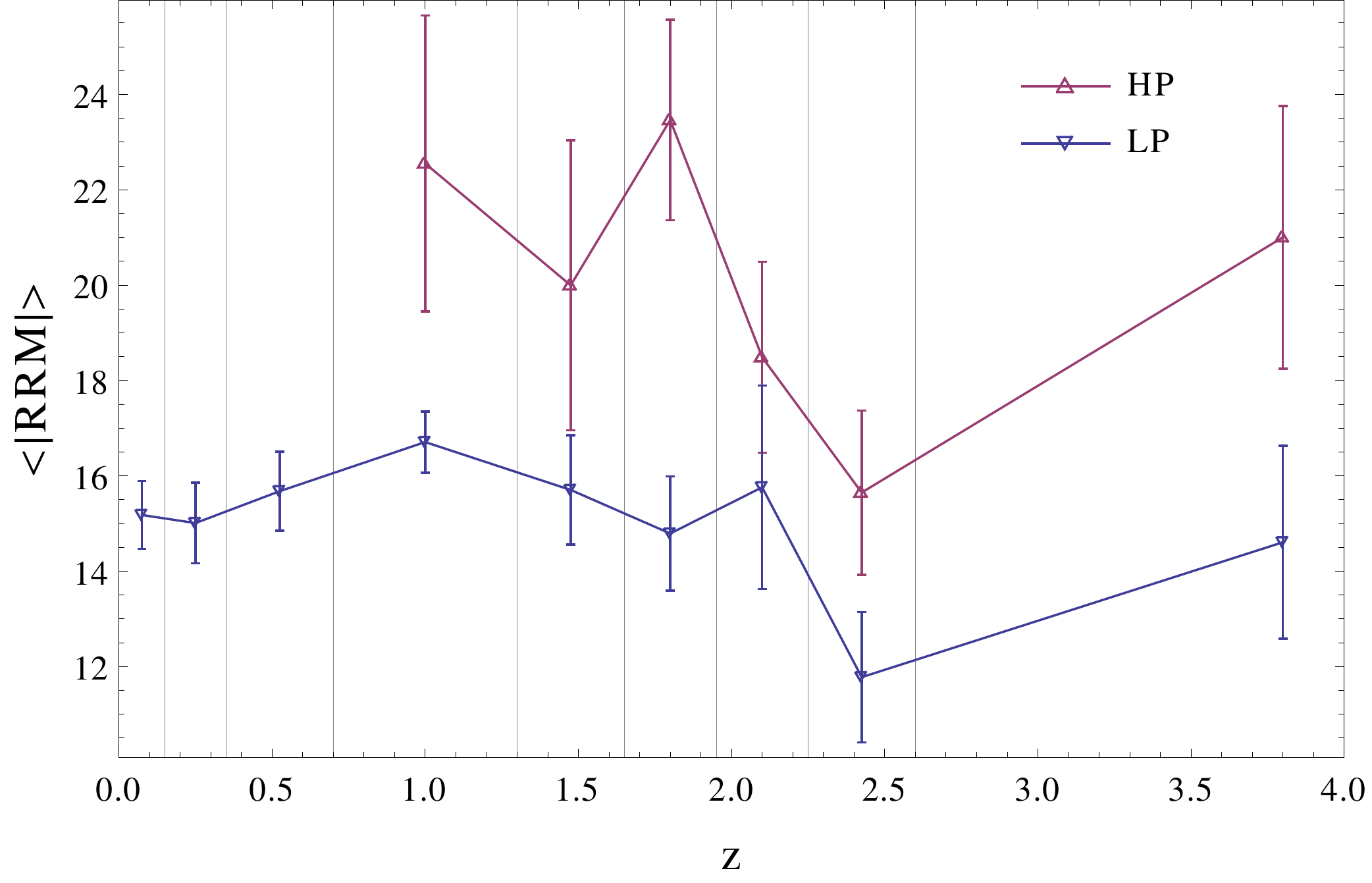}
\end{center}
\caption{Mean $\aRMinavg$ in z-bins for \lp and \hp sets separately for our set of sources with manually calculated luminosity.  All quantities are expressed in $\text{rad}/\text{m}^2$.}
\label{fig:mean20OLD}
\end{figure}

\begin{figure}
\begin{center}
\includegraphics[width=0.48\textwidth]{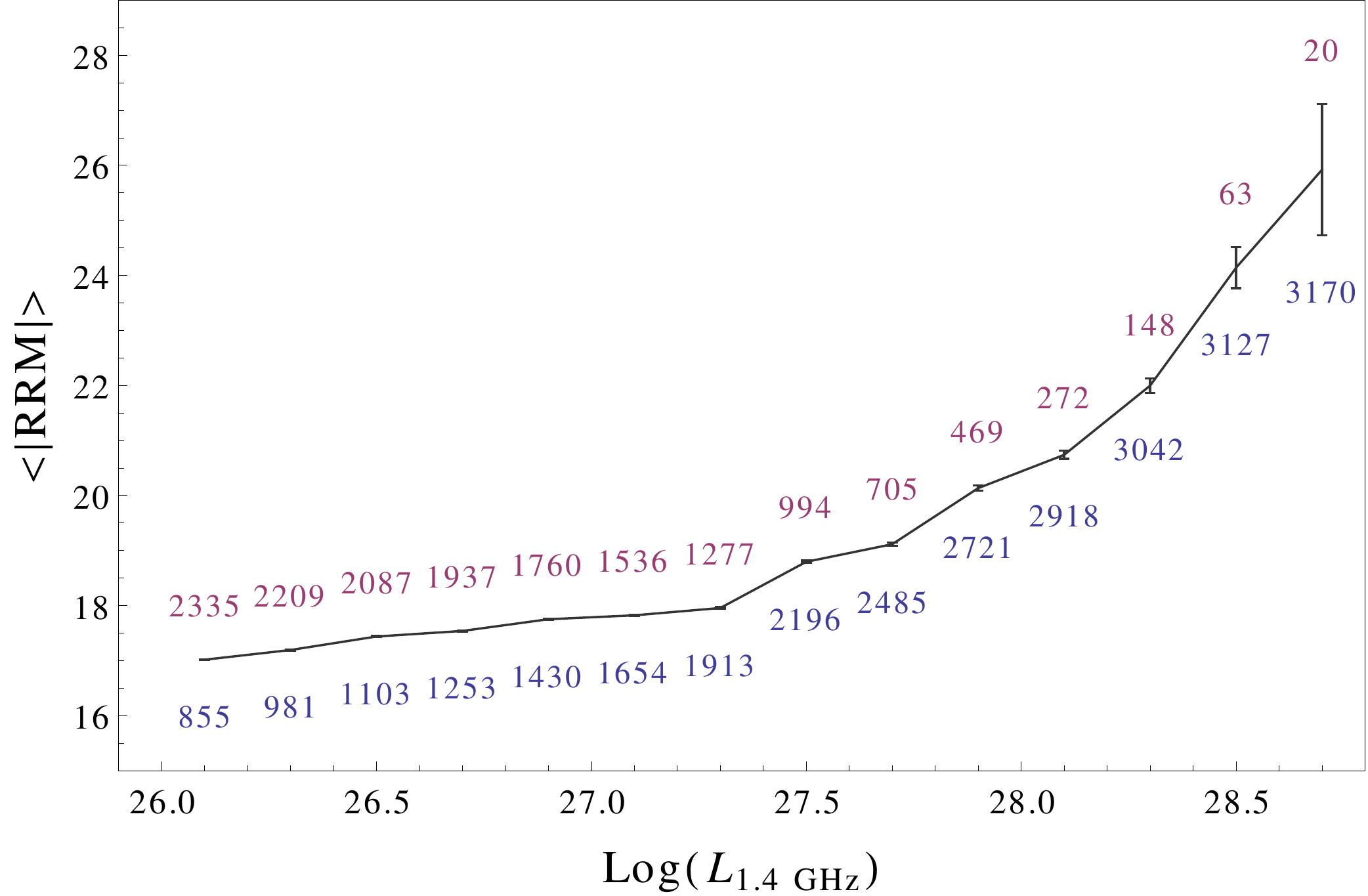}
\end{center}
\caption{Overall mean $|\RMin|$ for the \hp set plotted 
against luminosity cut.  The values above and below the curve are 
the number of events at given power threshold in the \hp 
and \lp sets, respectively, as defined by the cut.  All quantities are expressed in $\text{rad}/\text{m}^2$.}\label{fig:hzhpOLD}
\end{figure}

After calculating the luminosities again using Eq.~\ref{lumi} we can perform the analysis as we did in the main text.  We report here only the most relevant plots, that is, Fig.~\ref{fig:mean20OLD} with the redshift-binned $|\RMin|$ for the split set, (analogue to Fig.~\ref{fig:mean20j}), and Fig.~\ref{fig:hzhpOLD} showing the luminosity-RM correlation (analogue to Fig.~\ref{fig:hzhp}).

The features which we observed in the main text appear here unchanged---in fact, they are even more prominent: there is a positive correlation between the observed residual RM and the luminosity.

\newpage
\bibliographystyle{mn2e}
\bibliography{nvss}


\end{document}